\documentclass[12pt]{article}
\usepackage{graphicx}
\usepackage{subfigure}
\usepackage{units}

\newcommand\pubnumber{DPF2015-288}
\newcommand\pubdate{\today}

\def\umn{
University of Minnesota, Minneapolis, MN 55126}

\def\Title#1{\begin{center} {\Large #1 } \end{center}}
\def\Author#1{\begin{center}{ \sc #1} \end{center}}
\def\Address#1{\begin{center}{ \it #1} \end{center}}

\newcommand\pubblock{\rightline{\begin{tabular}{l} \pubnumber\\
         \pubdate  \end{tabular}}}
\newenvironment{Abstract}{\begin{quotation}  }{\end{quotation}}
\newenvironment{Presented}{\begin{quotation} \begin{center} 
             PRESENTED AT\end{center}\bigskip 
      \begin{center}\begin{large}}{\end{large}\end{center} \end{quotation}}
\def\Acknowledgments{\bigskip  \bigskip \begin{center} \begin{large}
             \bf ACKNOWLEDGMENTS \end{large}\end{center}}

\def\beq{\begin{equation}}
\def\eeq#1{\label{#1}\end{equation}}
\def\eeqn{\end{equation}}

\def\beqa{\begin{eqnarray}}
\def\eeqa#1{\label{#1}\end{eqnarray}}
\def\eeqan{\end{eqnarray}}

\let\bar=\overbar

\def\Dslash{\not{\hbox{\kern-4pt $D$}}}
\def\dslash{\not{\hbox{\kern-2pt $\del$}}}

\def\msb{{\bar{\ssstyle M \kern -1pt S}}}

\begin{document}
\begin{titlepage}
\pubblock

\vfill
\Title{The CMS Beam Halo Monitor Detector System}
\vfill
\Author{ Kelly Stifter\\ On behalf of the CMS collaboration}
\Address{\umn}
\vfill
\begin{Abstract}
A new Beam Halo Monitor (BHM) detector system has been installed in the CMS cavern to measure the machine-induced background (MIB) from the LHC. This background originates from interactions of the LHC beam halo with the final set of collimators before the CMS experiment and from beam gas interactions. The BHM detector uses the directional nature of Cherenkov radiation and event timing to select particles coming from the direction of the beam and to suppress those originating from the interaction point. It consists of 40 quartz rods, placed on each side of the CMS detector, coupled to UV sensitive PMTs. For each bunch crossing the PMT signal is digitized by a charge integrating ASIC and the arrival time of the signal is recorded. The data are processed in real time to yield a precise measurement of per-bunch-crossing background rate. This measurement is made available to CMS and the LHC, to provide real-time feedback on the beam quality and to improve the efficiency of data taking. Here, I present the detector system and first results obtained in Run II.
\end{Abstract}
\vfill
\begin{Presented}
DPF 2015\\
The Meeting of the American Physical Society\\
Division of Particles and Fields\\
Ann Arbor, Michigan, August 4--8, 2015\\
\end{Presented}
\vfill
\end{titlepage}
\def\thefootnote{\fnsymbol{footnote}}
\setcounter{footnote}{0}

\section{Introduction}

The new Beam Halo Monitor (BHM) is a novel detector that provides an online, bunch-by-bunch, Machine Induced Background (MIB) rate at large radii in the CMS experiment at the LHC. 

Machine Induced Background is produced through several interactions, but mainly through protons scattering off LHC collimators. MIB particles often arrive at a high radius, and can increase the fake trigger rate of muon chambers or leave energy deposits in calorimeters. The rate of MIB particles also gives information about the condition of the condition of the beam and can alert the LHC to any instabilities in the beam. 

The detector units, comprised of quartz bars read out by photomultiplier tubes, take advantage of precision timing and the directional nature of Cherenkov radiation in order to distinguish the low rate of MIB particles from the large flux of collision products.

\section{Detector system}

\subsection{Concept}

In order to separate the order of \unit[1]{$\textrm{Hz}/\textrm{cm}^{2}$} MIB signal from the \unit[$10^4$]{$\textrm{Hz}/\textrm{cm}^{2}$} collision (PP) product background, BHM uses a direction-sensitive detector and precision timing.

\begin{figure}[htb]
	\centering
	\includegraphics[width=.7\textwidth]{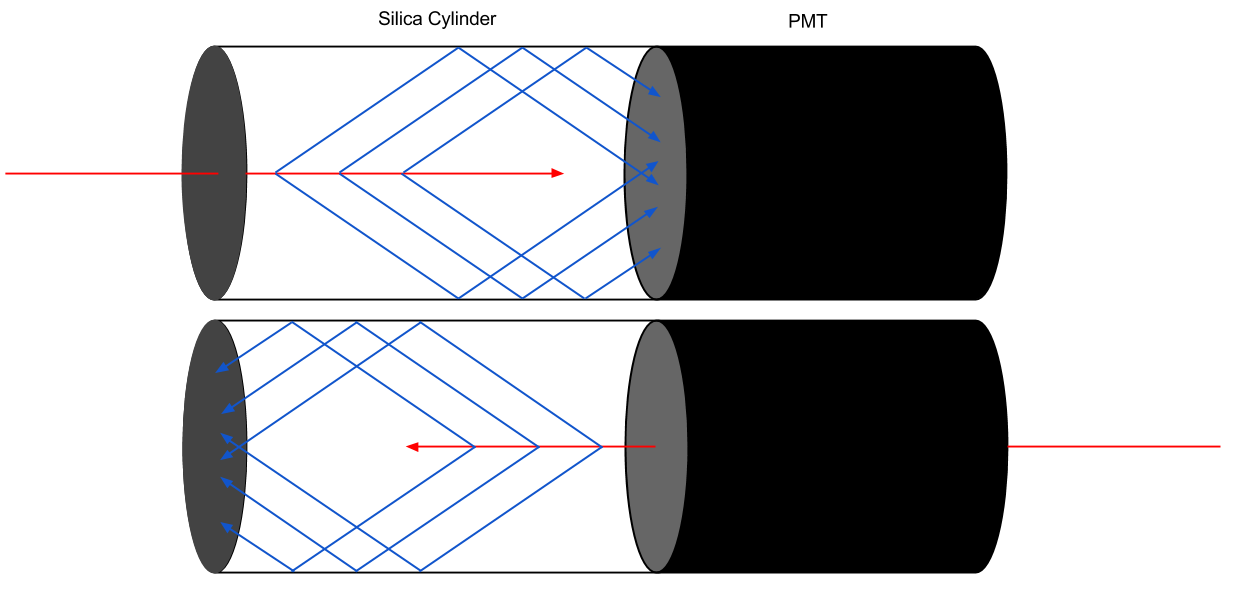}
	\caption{\textit{Top:} A MIB muon arrives with the incoming beam and produces Cherenkov radiation, which is detected by the PMT. \textit{Bottom:} A PP product arrives in the opposite direction, and the Cherenkov radiation that is produced is absorbed by black paint at the end of the cylinder.}
	\label{fig:concept}
\end{figure}

The base of a BHM detector unit is a fused silica cylinder optically coupled to a photomultiplier tube (PMT). The detector units are oriented in such a way that when a MIB particle arrives with the incoming beam and produces Cherenkov radiation in the quartz, the light propagates forward and is collected by the PMT. On the other hand, collision products arrive in the opposite direction. While they still produce Cherenkov radiation, the resulting photons propagate in the opposite direction and are absorbed by black paint that is applied to the other end of the cylinder. Consequently, signals from backward particles are much smaller than those from forward particles. Both of these processes are shown in Fig.~\ref{fig:concept}.

This concept was tested with particle beams at CERN in 2012 \cite{IBIC} and DESY in 2014 \cite{NThesis}, and it was shown  that by implementing a charge amplitude cut on all signals, it is possible to suppress signals from backward particles by a factor of $10^{4}$ while maintaining an efficiency close to 100$\%$ for MIB signals.

In addition to using the directional nature of the MIB and PP particles, the BHM detector uses precise timing to separate these two types of events. Since the bunches from the LHC collide in CMS at the interaction point every \unit[25]{ns}, and all products are traveling with the beams at the velocity of light, it is possible to select a location along the beam line where the time difference between the arrival of MIB and PP particles is maximized. The BHM detectors were placed at one of these locations on each side of CMS in order to take advantage of this property.

\subsection{Detector units}

The Cherenkov medium used in the detector units is a UV-transmissive, radiation hard, \unit[10]{cm} cylinder of SQ0 synthetic fused silica, which is \unit[5.2]{cm} in diameter. The cylinder is optically coupled to a Hamamatsu R2059 photomultiplier tube of the same diameter using a silicon disk. This PMT was chosen for its sensitivity to UV light. The front of the cylinder is painted black in order to absorb the Cherenkov radiation that propagates in the opposite direction.

\begin{figure}[htb]
	\begin{center}	
	\subfigure[Detector Units]
	{
		\label{fig:units}
		\includegraphics[width=0.47\textwidth]{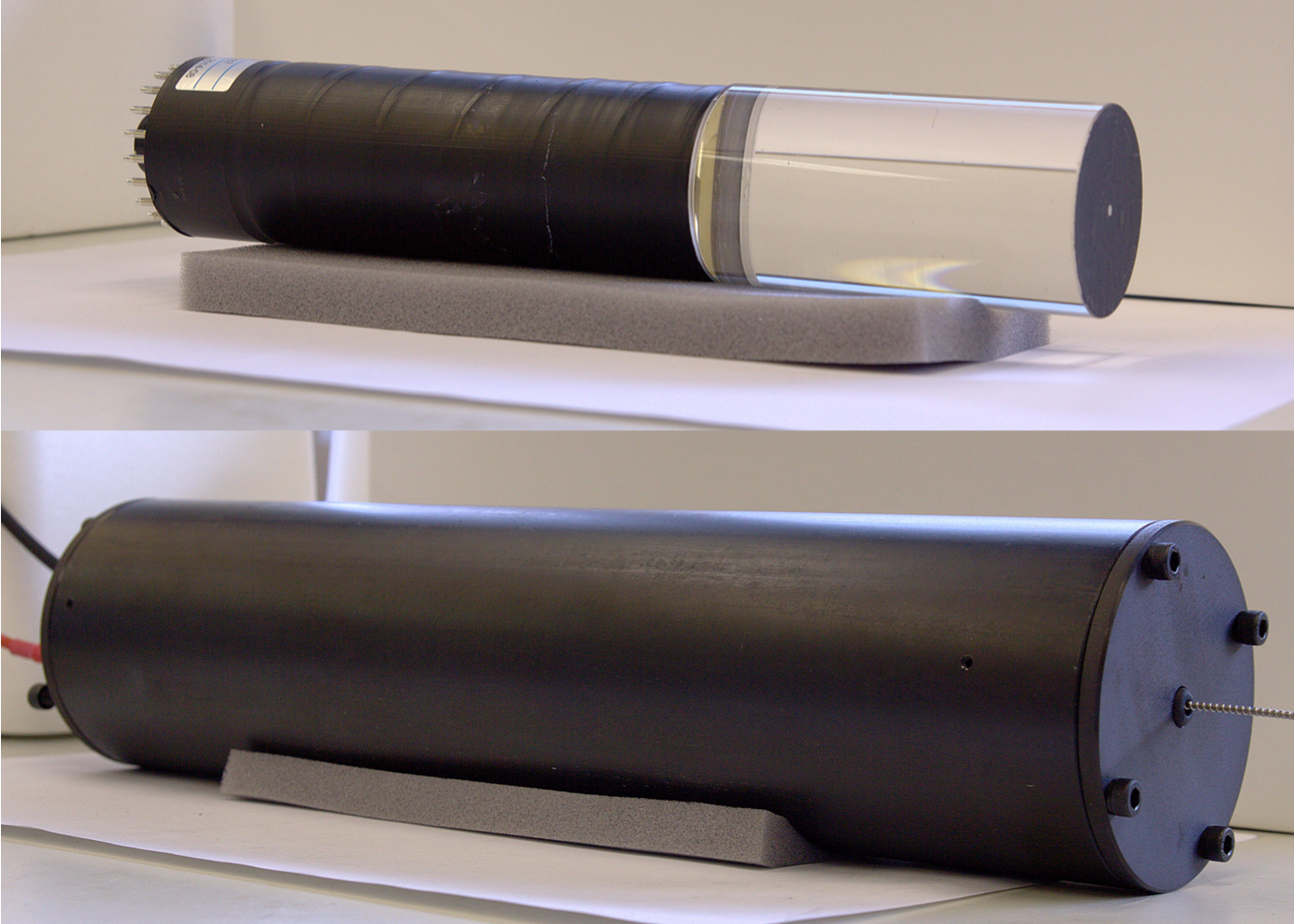}
	}
	\subfigure[Unit Orientation]
	{
		\label{fig:orientation}
		\includegraphics[width=0.47\textwidth]{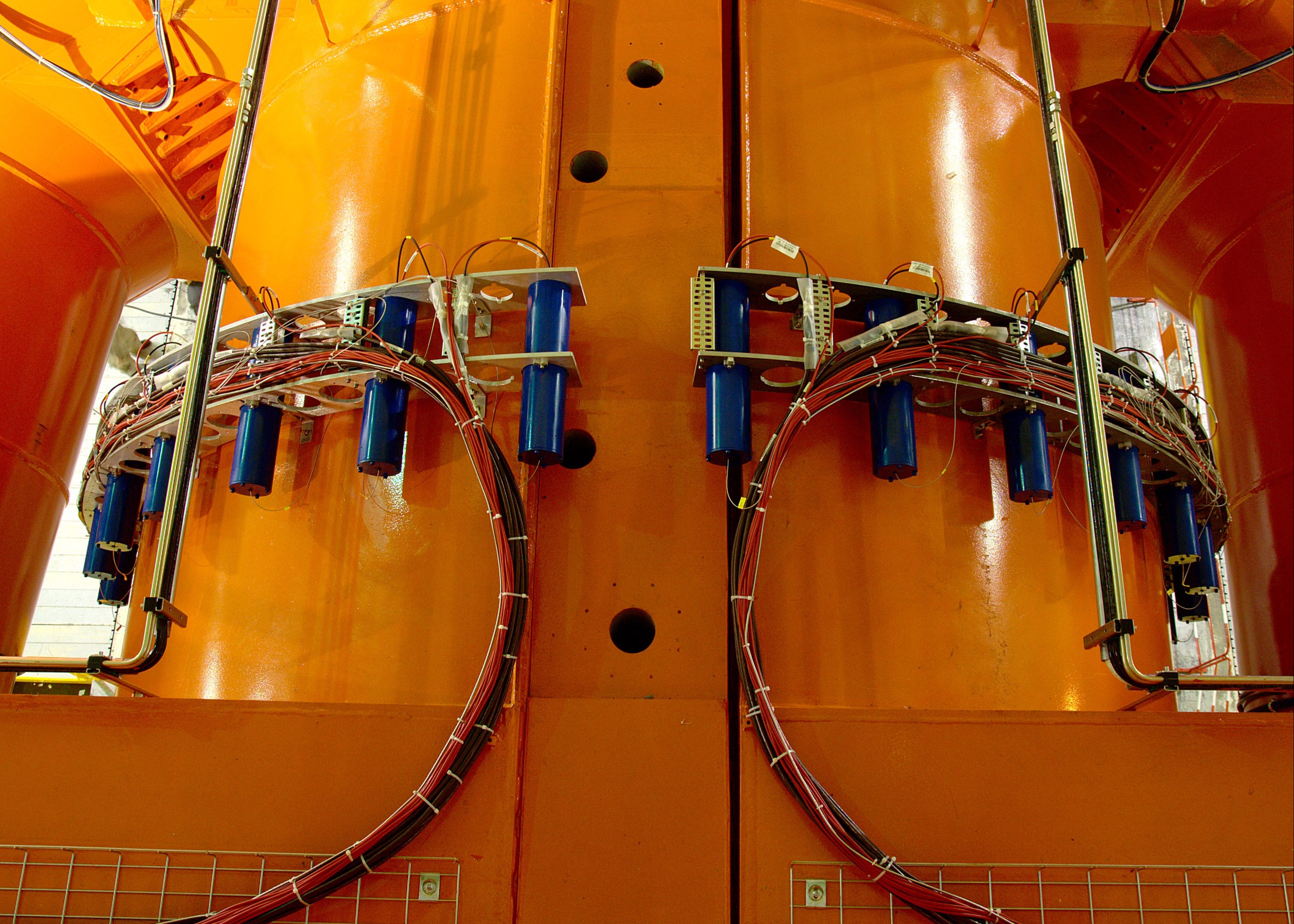}
	}
	\caption{\textit{Left:} The base of a detector unit, and the final layer of \unit[1]{cm} iron shielding. \textit{Right:} Azimuthal distribution of BHM detector units around CMS magnetic shielding. The units point away from the bulk of CMS, in order to detect incoming particles.}
	\end{center}
\end{figure}

In order to protect the sensitive PMT from any residual magnetic field from the CMS magnet, three layers of magnetic shielding were used. The first is a layer of Permalloy. In addition to shielding, it serves as the mechanical support for the PMT and quartz bar. The second is a layer of mu-metal. A \unit[1]{cm} thick cylinder of iron is used as the third and final layer, as shown in Fig.~\ref{fig:units}. This layer also absorbs much of the large flux of low-energy particles present in the cavern.

Twenty of these units are installed on each side of CMS, \unit[20.65]{m} from the interaction point, which is outside the bulk of the detector. This location was selected for a number of different reasons, including a high MIB flux compared to collision products, a small residual magnetic field from the CMS solenoid, a low overall radiation dose, and easy access for installation, cabling, and services.

The units are oriented parallel to the beam line, and are distributed azimuthally, at a radius of \unit[1.8]{m} from the beam line. The detectors located on either end measure the background for the incoming beam, with a total acceptance of \unit[424]{cm$^2$}.

\subsection{Read-out system}

The backbone of BHM electronics comes from the HCAL Phase 1 Upgrade \cite{HCAL}. The analog signal from each PMT is read out by front-end digitizer cards, called HF Readout Modules \cite{QIE}, and then sent, via \unit[5]{Gbps} asynchronous optical link, to the back-end histogramming unit, called a MicroTCA HCAL Trigger and Readout unit ($\mu$HTR) \cite{uHTR}. From there, the data is sent out to software, which is a part of xDAQ-based architecture called BRIL-DAQ \cite{xdaq}.

The data comes as eight-bit charge information and six-bit timing information for each detector unit during each bunch crossing. A per-channel amplitude cut is performed, and then events passing the threshold are binned in an occupancy histogram by their bunch crossing and the timing information, which is mapped into four TDC bins. The current mapping sets one bin to the time a MIB signal is expected, another to the time a collision product is expected, and the other two for early and late hits. The histograms are then read out by the software, where a flux is calculated and published to CMS and the LHC every \unit[23]{s}.

This system serves as a first demonstration of full functionality of all electronics components, including the pre-production front-end cards and the high-speed \unit[5]{Gbps} asynchronous link used.

\subsection{Calibration system}

A calibration system \cite{NThesis} was installed in order to assess the detector performance over time as radiation damage and aging effects start to become significant. The calibration system is designed to distribute light pulses of known amplitude to all detector units in order to measure variation in PMT response.

\section{Commissioning results}

The first BHM commissioning results were recorded during the LHC “beam splash” events in April 2015. During a splash event, one bunch of protons is directed into the tertiary collimators, which are located immediately before the experimental cavern. This produces a large flux of secondary particles, which travel through CMS in the same direction, at the same time.

\begin{figure}[htb]
	\centering
	\includegraphics[width=.6\textwidth]{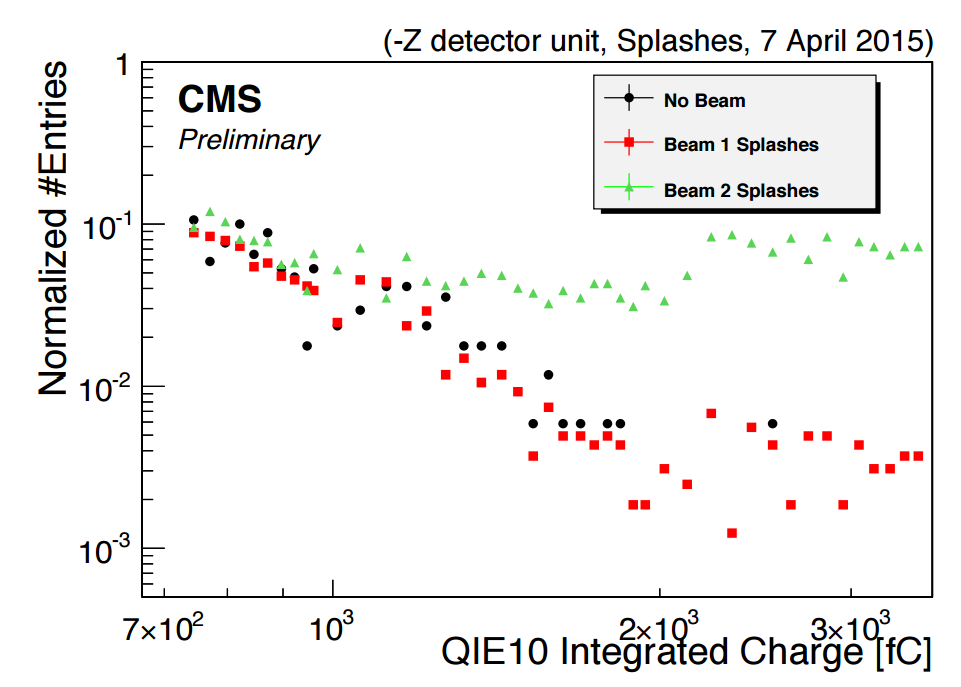}
	\caption{Amplitude data from a beam 2 detector during beam 1 splashes, beam 2 splashes, and no beam.}
	\label{fig:splashes}
\end{figure}

Fig.~\ref{fig:splashes} shows the distribution of the charge recorded in a single detector unit that measures the halo for Beam 2. An increased fraction of high-amplitude events (green) was seen during the Beam 2 splashes, whereas the response to the Beam 1 splashes (red) is similar to the one with no beam (black). This shows that while the unit saw many high-amplitude forward particles, no high-amplitude backward particles were observed. This served as the first indication of the directional sensitivity of the BHM detectors.

\begin{figure}[htb]
	\centering
	\includegraphics[width=.6\textwidth]{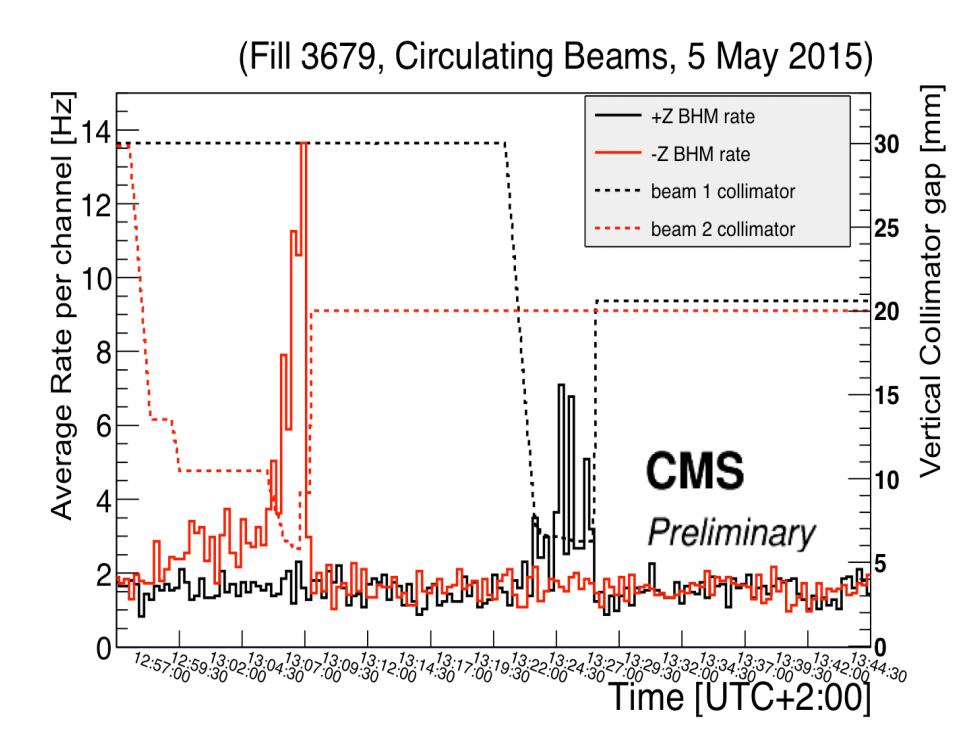}
	\caption{MIB rate shows distinct correlation to the collimator gap.}
	\label{fig:collimatorscan}
\end{figure}

Narrow collimator aperture is expected to be a large source of MIB particles. The sensitivity of BHM to this movement is demonstrated in Fig.~\ref{fig:collimatorscan}. The dashed lines show the collimator aperture for the two opposing beams, while the solid lines show the average background rate in the BHM detectors that look in opposing directions. A sharp increase in BHM rate occurs for each end when the collimator aperture of the respective beam is reduced, while the rate in the opposite detectors remains the same.

\begin{figure}[htb]
	\begin{center}	
	\subfigure[Occupancy Histogram]
	{
		\label{fig:occupancy}
		\includegraphics[width=0.4\textwidth]{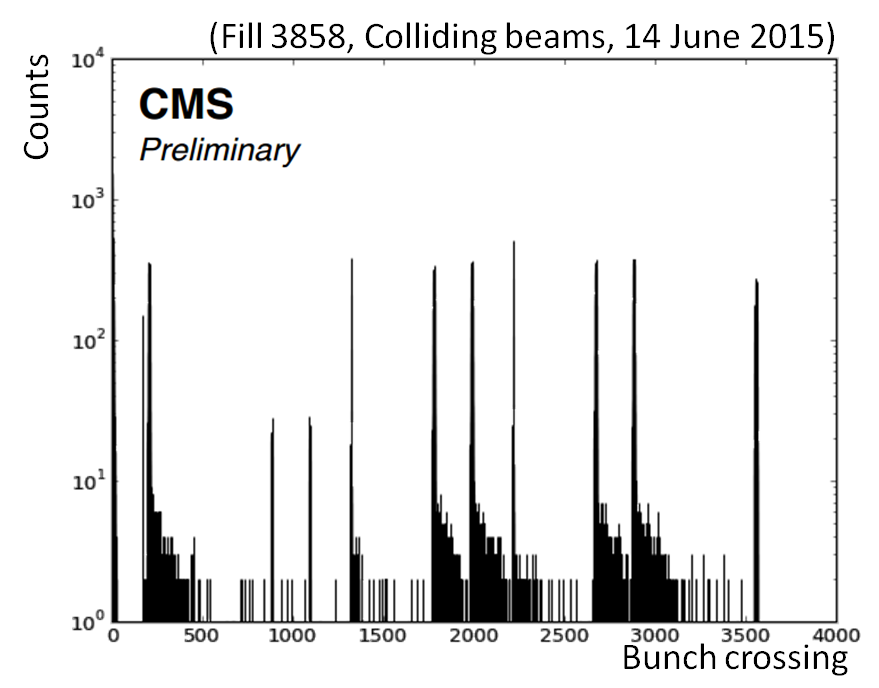}
	}
	\subfigure[Single Bunch Train]
	{
		\label{fig:occupancy_zoom}
		\includegraphics[width=0.53\textwidth]{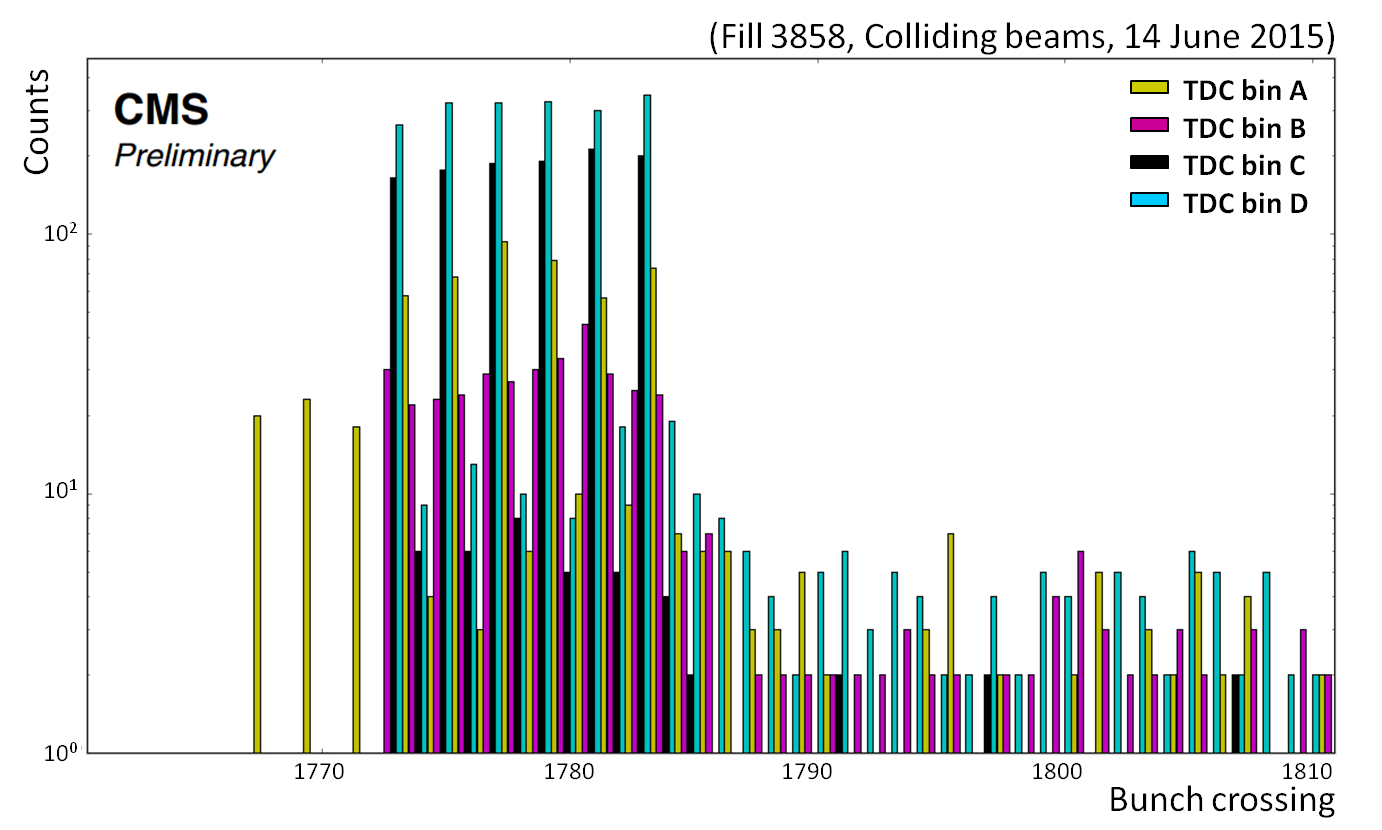}
	}
	\caption{\textit{Left:} Summed occupancy histograms for fill 3858 from a single detector unit. \textit{Right:} A closer look at a bunch train in fill 3858 from a single detector unit. Please note that the given bunch crossing number was not yet aligned with the global bunch crossing number at this stage of commissioning.}
	\end{center}
\end{figure}

As the commissioning process continued, data from stable collisions were collected with low signal amplitude thresholds. The details of the fill structure can be seen as changes in the Machine Induced Background (MIB) rate for a selected detector unit, as shown in fig.~\ref{fig:occupancy}. In this fill, there were six bunch trains, two single bunches, and two probe bunches. Each bunch train contained six colliding bunches spaced at \unit[50]{ns}, and is followed by a tail of hits, the timing of which is consistent with the expected albedo in the CMS cavern. The two single bunches have the same amplitude as the bunch trains, and are similarly followed by an albedo tail. The two probe bunches were of much lower intensity, which account for their smaller amplitude and lack of albedo tail.

A closer look at a single bunch train reveals more details, as can be seen in Fig.~\ref{fig:occupancy_zoom}. The first three filled bins are purely Machine Induced Background (MIB) particles. The first collision products are seen approximately 5.5 bunch crossings later, as expected. This pattern then continues for six subsequent bunches. Note that all MIB events appear in TDC bin 1, which represents the same \unit[6.25]{ns} window within each bunch. Alternatively, the particles from collisions come mainly in different TDC bins - either 0 or 3. This indicates that BHM will be able to successfully distinguish MIB and PP particles based on the sub-bunch crossing timing information.

\section{Summary}

The new BHM detector was successfully installed in CMS during LS1. Preliminary results indicate that the detector is sensitive to MIB particles produced through interactions with collimators, the directional aspect of the detector units can be used to discriminate MIB particles and PP products, and these signals can be further differentiated through the use of sub-bunch crossing timing.

\Acknowledgments
These activities are the collaborative work of a number of physicists, engineers, and technicians from several institutions, including CERN, University of Minnesota, INFN Bologna, and National Technical
University of Athens. Worthy of special note are the contributions of A. E. Dabrowski, J. Mans, S. Orfaneli, R. Rusack, and N. Tosi.


\begin{thebibliography}{99}

\bibitem{IBIC}
S. Orfanelli, et al. Design of a novel Cherenkov detector system for Machine Induced Background monitoring in the CMS cavern. IBIC2013, (Oxford, UK, 2013) 33–36.
\bibitem{NThesis}
N Tosi. The new Beam Halo Monitor for the CMS experiment at the LHC. PhD thesis, (Alma Mater Studiorum Universit\`a di Bologna, 2015).
\bibitem{QIE}
A. Baumbaugh et al. QIE10: a new front-end custom integrated circuit for high rate experiments. JINST, 9(01), (2014).
\bibitem{uHTR}
J Mans. HCAL uHTR Specifications and Operational Documentation. CMS-DOC-12306-v12. (2014).
\bibitem{HCAL}
J. Mans, et al. CMS Technical Design Report for the Phase 1 Upgrade of the Hadron Calorimeter. Technical Report CERN-LHCC-2012-015, CMS-TDR-10 CERN (Geneva, Switzerland, 2012).
\bibitem{xdaq}
V. Brigljevic et al. Using XDAQ in application scenarios of the CMS experiment. FERMILAB-CONF-03-293, CHEP-2003-MOGT008, (2003).

\end{thebibliography}
\end{document}